\documentclass[12pt]{iopart}
\usepackage{iopams} 
\usepackage{hyperref,graphicx}

\def\Dslash{\,\,{\raise.15ex\hbox{/}\mkern-12mu D}}
\def\Dbarslash{\,\,{\raise.15ex\hbox{/}\mkern-12mu {\bar D}}}
\def\delslash{\,\,{\raise.15ex\hbox{/}\mkern-9mu \partial}}
\def\delbarslash{\,\,{\raise.15ex\hbox{/}\mkern-9mu {\bar\partial}}}
\def\pslash{\,\,{\raise.15ex\hbox{/}\mkern-9mu p}}
\def\calDslash{\,\,{\raise.15ex\hbox{/}\mkern-12mu {\cal D}}}

\newcommand{\mO}{\mathcal{O}}

\begin{document}

\title{Quantum Information Processing and Relativistic Quantum Fields}

\author{Dionigi M. T. Benincasa${}^1$, Leron Borsten${}^1$, Michel Buck${}^1$,
Fay Dowker${}^{1,2}$}
 \address{${}^1$Blackett Laboratory, Imperial College, London, SW7 2AZ, U.K.}
 \address{${}^2$Institute for Quantum Computing, University of Waterloo, ON, N2L 3G1, Canada}
\begin{abstract}

It is shown that an ideal measurement of a one--particle wave packet state of a relativistic 
quantum field in Minkowski spacetime enables superluminal signalling. 
The result holds for a measurement that takes place over an intervention region in 
spacetime whose extent in time in some frame is longer than the light crossing time of the 
packet in that frame. Moreover, these results are shown to apply  not only to ideal measurements but also to  unitary transformations that rotate  two orthogonal one--particle states into each other. In light of these observations, possible restrictions on the allowed types of intervention are considered. A more physical approach to such questions is to construct explicit models of the interventions as interactions between the field and other quantum systems such as detectors. The prototypical Unruh-DeWitt detector couples to the field operator itself and so most likely respects relativistic causality. On the other hand, detector models which couple to a finite set of frequencies of field modes are shown to lead to superluminal signalling. Such detectors do, however, provide successful phenomenological models of atom-qubits interacting with quantum fields in a cavity but are valid only on time scales 
many orders of magnitude larger than the light crossing time of the cavity.

\end{abstract}

\maketitle
\section{Introduction}

In a typical quantum information processing scheme classical agents 
use a quantum system
to encode, process and communicate information (classical bits).  
A demand to be made of a description of a quantum information
processing (QIP) protocol, 
 then, is: show me the bits. In other words it must be understood how 
the classical bits are encoded physically in the quantum system and how they
are read out.  A qubit is a two-dimensional quantum system which can be a building block of the 
 total quantum system which is used to perform the information processing 
 task. In a typical protocol  external agents feed bits into the system
 by preparing some of the qubits in one quantum state or another and read bits
out at the end by 
 making measurements on some (other) of the qubits. Intermediate steps 
 may involve the performance of unitary transformations and other sorts of
operations on the qubits by the agents. So a related demand
 is: show me the physical qubits. This is particularly important in 
 relation to entanglement. After all, given {\textit{any}} 
 pure state, $| \psi\rangle$, in a 4-dimensional Hilbert space, 
 there is an isomorphism from that Hilbert 
 space to a tensor product of two 2-dimensional spaces -- two `qubits' --  in which 
 $| \psi\rangle$ is mapped to a maximally entangled state as expressed in the product basis. 
These `qubits'  will generally be useless for QIP because they will not correspond 
 to any physically accessible two-dimensional systems that can be manipulated 
 by external agents. Thus, in assessing the relevance of calculations of measures of 
 entanglement between qubits for QIP, it is necessary to determine 
 whether and how external agents can intervene upon them. 
 
The satisfaction of these demands to identify how bits are encoded into 
and read out from useful qubits 
and how those qubits can be operated upon physically
becomes particularly challenging 
when investigating QIP in a relativistic spacetime, taking 
into account the
locations in spacetime of the actions of the external agents on the 
quantum system.  An obvious   framework 
for such  investigations is relativistic quantum field theory. Here, progress is hampered by the lack of a 
universally applicable rule for calculating the probabilities of 
the outcomes of ideal measurements on a relativistic quantum field 
in a collection of spacetime regions. Indeed,
a straightforward relativistic generalisation of the 
non-relativistic formula for these probabilities
leads to superluminal signalling \cite{Sorkin:1993gg}. 

We review this generalised rule. We work in the Heisenberg (Interaction) Picture.  Let ${\mathcal{O}}_i$, $i = 1,\dots n$ be a 
collection of regions in a globally hyperbolic spacetime. 
 Let $\preceq$ be a relation on the regions
defined by 
$\mO_j \preceq \mO_k$ iff some point in $\mO_j$ is in the causal past of some point in $\mO_k$. 
The transitive closure of this relation is taken and the resulting relation denoted by the
same symbol $\preceq$.
If the resulting relation is acyclic -- {\textit{i.e.}}  $\mO_j \preceq \mO_k$ and $\mO_k \preceq \mO_j$ implies $j=k$ -- then it is a partial order
and there exists a linear ordering of the regions which is compatible with the partial order. 
In other words there is an assignment of labels $i = 1,\dots n$ to the regions such that 
$\mO_j \preceq \mO_k $ implies $j\le k$. We assume from now on that the regions
and the labelling
satisfy this condition.  To each region $\mO_i$ there corresponds an algebra of observables.
Consider, then, for each $i$, the measurement of an observable 
$A_i$ in the algebra associated to region $\mO_i$ and consider a particular possible outcome of that 
measurement corresponding to a (Heisenberg) projection operator, $P_i$ say, onto the
relevant eigenspace of the observable.  
The probability of obtaining those particular outcomes to the sequence of measurements
 in the regions $\mO_i$ is proposed to be \cite{Sorkin:1993gg}
  \begin{equation}\label{eq:probs}
 {\textrm{Tr}}\left(P_n \dots P_1 \rho P_1 \dots P_n\right) 
 \end{equation}
where $\rho$ is the initial state.\footnote{If one wants to interpret the rule as corresponding to an 
effect on the quantum state of the field in the interaction picture, it would be that for each intervention 
region $\mO_i$ in turn -- in the linear order
defined by the labels, $i = 1,\dots n$ --  the quantum state collapses along the boundary of the causal past of the region $\cup_{k = 1}^i \mO_k$ \cite{Hellwig:1970st}.}

One of the assumptions inherent in Sorkin's proposal for the probability (\ref{eq:probs}) 
is that the measurement of
$A_i$ is achieved by means of procedures that take place entirely within $\mO_i$.
Since, if the intervention were located anywhere outside $\mO_i$ that would
imply further restrictions on the possible locations of the other regions $\mO_j$: they 
shouldn't overlap with the location of the operation of measuring $A_i$ for example.
We will refer to  region $\mO_i$ therefore as the
 ``intervention region'' in which the external agents act upon 
the field to effect the measurement 
of $A_i$.  Note that the intervention region, $\mO_i$ 
can be larger but not (presumably) 
smaller than any minimal region in which observable $A_i$ is defined. 
For example, $A_i$ could belong to the algebra of observables associated 
to a region $\mathcal{N}_i$ which is a proper subset of $\mO_i$, but the 
 the measurement intervention could
be being made throughout $\mO_i$. 

In the nonrelativistic theory, an ideal measurement is ``idealized" in more than 
one way. It is assumed to result in the precise collapse of the state into an
eigenstate of the observable and it is assumed to be achievable by an 
external intervention whose extent in time is negligible compared to the
characteristic  timescale of the unitary evolution of the system being intervened upon.
Then, a mathematical model of  a ``Von Neumann" measurement in which a quantal
detector is coupled formally to the nonrelativistic system can be constructed which comes
arbitarily close to achieving such an ideal measurement.
In adopting (\ref{eq:probs}) as the relativistic analogue of the nonrelativistic formula, 
the ``collapse to an eigenstate'' aspect of an ideal measurement is being preserved
but the ``short time extent compared to evolution of system'' aspect is clearly not 
being assumed from the outset -- there is no restriction on the temporal
extent of the regions, a priori. The results of Sorkin and those
reported here can be interpreted as showing that for certain, seemingly physically 
meaningful ``observables'' 
of a relativistic quantum field theory, constructing a realistic Von Neumann 
type measurement model is impossible. 

\section{Impossible measurements of wave packets}
\begin{figure}
\includegraphics[width=\textwidth]{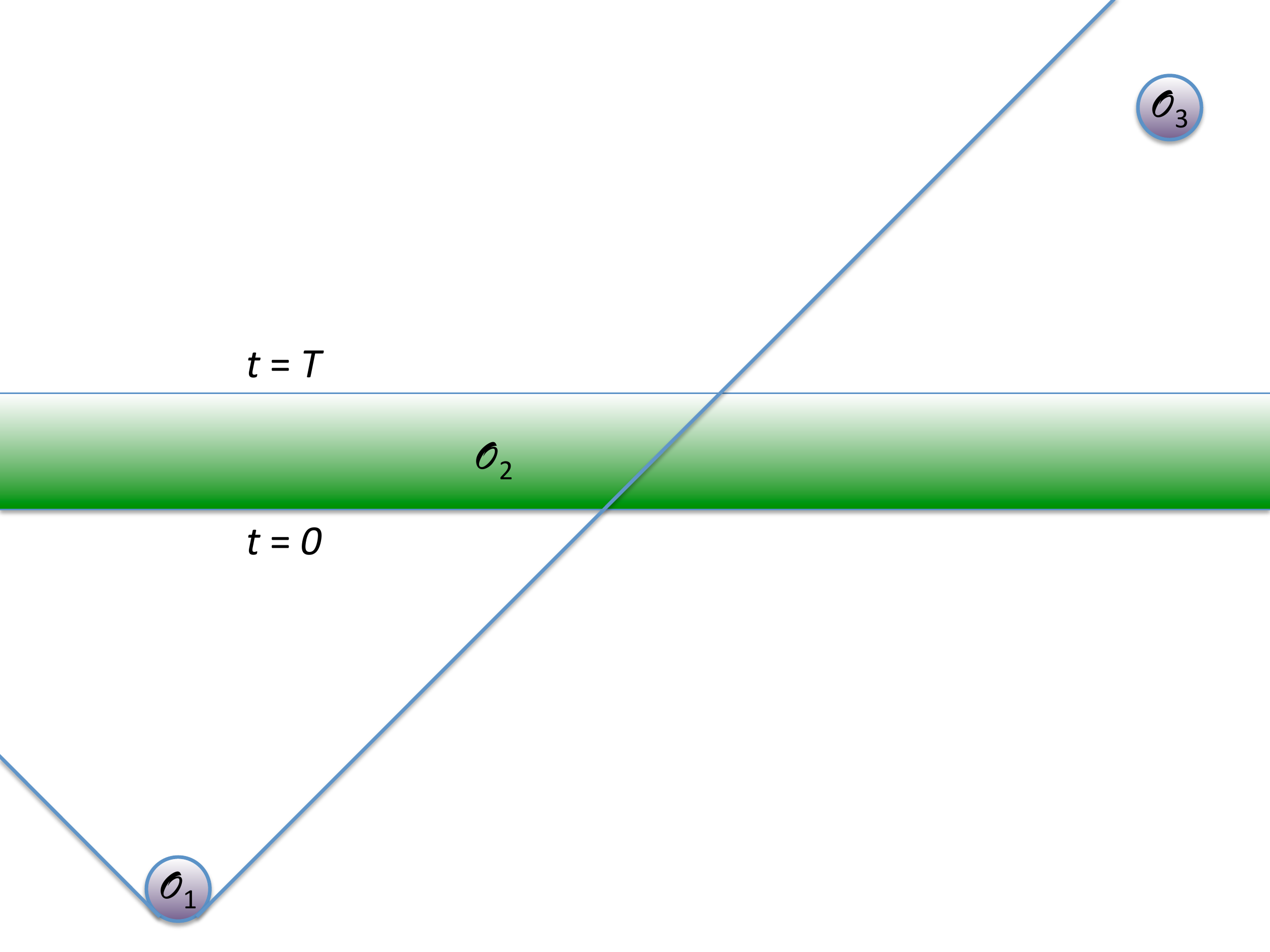}
\caption{The three intervention regions. In region 1 (a small open ball around spacetime point $X$) we apply a local unitary transformation $e^{i\lambda \hat{\phi}(X)}$. In region 3 (a small open ball around spacetime point $Y$) we measure $\hat{\phi}(Y)$. In region 2 (a slab of spacetime between $t=0$ and $t=T>0$), we consider both ideal measurements of one-particle states and unitary rotations between distinct one-particle states.}
\label{fig:setup}
\end{figure}
Henceforth, we will explicitly consider $(d+1)$-dimensional Minkowski space
with a mostly plus metric convention,
and a free massless scalar field $\hat\phi(x)$, though much of what
follows will apply with little or no modification for free massive fields and 
the general conclusions are relevant in globally hyperbolic spacetimes. 
Using the formula (\ref{eq:probs}), the following protocol will generally
result in superluminal signalling \cite{Sorkin:1993gg}. There are three intervention regions:
$\mO_1$ is a (small open ball around) spacetime point $X^\mu$ where $X^0 \le0$, 
$\mO_2$ a slab of spacetime between $t=0$ and $t=T>0$ and $\mO_3$ 
is a (small open ball around) spacetime point $Y^\mu$ spacelike to 
$X$ and with $Y^0 \ge T$. See \Fref{fig:setup}. 
The interventions in these regions are, respectively, 
a unitary transformation $e^{i \lambda \hat\phi(X)}$ where $\lambda$ is a
real number,  a
measurement of $B := | 1\rangle\langle 1 |$ where $| 1\rangle$ is a one particle state
and a measurement of 
$\hat\phi(Y)$.

We pause here to elucidate the meaning of the ``measurement of $B$". $B$ is a projection 
operator with eigenvalues  $1$ and $0$ and what is being measured -- putatively --
is whether the system is in (Heisenberg) state $| 1 \rangle$.  This observable, 
which corresponds to a time independent operator, $B$, in the Heisenberg picture, 
corresponds to a time dependent operator in the Schroedinger picture. For example, if the 
state corresponds to a one-particle wavepacket -- as we will consider in what follows -- the observable 
$B$ corresponds to different positions for the particle at different times as the wavepacket evolves. 
What is being assumed via formula (\ref{eq:probs}) 
applied to the protocol described above is that the measurement is achieved 
by interventions within the slab $\mO_2$ and the 
collapse to the eigenstate of $B$ has occurred by the hypersurface $t = T$ so that the 
measurement at $Y$ with $Y^0 \ge T$ can take place. 

According to  (\ref{eq:probs}), the expected value of the outcome of the
measurement of $\hat\phi(Y)$ depends on $\lambda$ indicating a superluminal signal
the strength of which is quantified by \cite{Sorkin:1993gg}
\begin{equation}
S(X,Y) := -{\textrm{Im}}\left(\psi(X)^* \psi(Y)\right)
\end{equation}
where $\psi(\xi) := \langle 0 | \hat\phi(\xi) | 1\rangle $ is the ``one--particle wave-function.''
$S(X,Y)$ is the derivative of the expected outcome  of the measurement of
$\hat\phi(Y)$ w.r.t. $\lambda$, at $\lambda=0$, which may be  obtained directly by setting $\alpha=0, \beta=1$ in equation (4) of \cite{Sorkin:1993gg}. 
Fixing spacetime point $X$ so that $\psi(X)$ is real and nonzero, we obtain ${\textrm{Im}}\left( \psi(Y)\right)$ as
a measure of the strength of the superluminal signal. 

It is straightforward to show that this can be nonzero 
when the state $| 1\rangle$ is a one particle state with a precise $d$-momentum, $\mathbf{k}$.
(Spatial $d$-vectors will be written in boldface.)
Moreover, in this case the wave-function satisfies $\psi(Y^\mu + \xi^\mu) = \psi(Y^\mu)$
where $\xi^\mu$ is any null vector 
proportional to the $(d+1)$-momentum, $k^\mu = 
(|\mathbf{k}|, \mathbf{k})$. So the superluminal signal remains no matter how large $T$, 
no matter how long the measurement-intervention takes.
We conclude that an ideal measurement 
of this single momentum one--particle state is impossible, if superluminal signalling is forbidden
(and assuming that the interventions at $X$ and $Y$ can be done). 
Such a result is not surprising, given the nonlocal character of a fixed momentum 
state: it is defined on an entire spacelike hypersurface. 
What might be more surprising 
is that similar conclusions can be drawn for ideal measurements of localised 
one--particle wave packet states as we will now show. 

Consider the Gaussian one--particle state
\begin{equation}
|1_d\rangle:=(\pi\sigma^2)^{-\frac{d}4}\int\,d^d{k}\,e^{-\frac{(\mathbf{k}-\mathbf{k}_0)^2}{2\sigma^2}}a^\dagger_{\mathbf k}|0\rangle
\label{eq:gauss1p}
\end{equation}
where $a^\dagger_{\mathbf{k}}$ is the
creation operator for a one particle state with 
$d$-momentum $\mathbf{k}$, 
$\mathbf k_0$ is the mean momentum, $\sigma$ is the spread in momentum space
and $|\mathbf k_0|\gg\sigma$. 

Consider first $d=1$ and let  $k_0^\mu = k_0(1,1)$ where $k_0 >0$, so the packet is moving
in the positive space direction. 
Then for 
any null vector $\xi^\mu \propto k_0^\mu$ we have:
\begin{equation}
\begin{array}{ll}
\psi(Y^\mu+\xi^\mu)&=\langle0|\hat\phi(Y^\mu+\xi^\mu)|1_1\rangle\\
&=(\pi\sigma^2)^{-\frac14}\int_{-\infty}^{\infty}\frac{ dk}{4\pi|k|}e^{-\frac{(k-k_0)^2}{2\sigma^2}}e^{ik_\mu(Y^\mu+\xi^\mu)}\,.
\end{array}
\end{equation}
To avoid the pole at the origin, let us modify the wavepacket slightly, 
allowing it to have support in momentum space
only for $k > \epsilon >0$ where $\epsilon$ is small.  
Then for all momenta contributing to the integral we have  $k_\mu \xi^\mu = 0$ and, hence, 
\begin{equation}
\psi(Y^\mu+\xi^\mu) =  \psi(Y^\mu)\,.
\end{equation}
In $1+1$ dimensions, when the packet has support only on momenta in the positive spatial 
direction, it holds its form, does not spread and the superluminal signal 
persists undiminished for arbitrary intervention time, $T$: 
$Y$ can be chosen in the support of the packet and such that 
$Y^0 = T$ and ${\textrm{Im}}(\psi(Y)) \ne 0$. Note that this result holds for
{\textit{any}} packet with support only 
on positive momenta -- the approximately Gaussian form is not necessary.

The quantum theory of a massless scalar field in $1+1$ dimensional Minkowski spacetime
is unphysical, not least because it
suffers from an infrared divergence, so we turn our attention to $3+1$ dimensions.
 The same conclusion obtained in $1+1$ dimensions about 
persistence of the signal holds in $3+1$ dimensions
if the wavepacket is completely spatially delocalised in the directions transverse to the direction of motion of the packet which is in the positive $z$-direction, say. In other words, 
the packet has support in momentum space on fixed values of the momentum in the 
$x$ and $y$ directions and a spread of positive values of the momentum in the $z$ direction.
 In the more physical case, when the packet is localised in all three spatial dimensions and described by the state $|1_3\rangle$ in~\eref{eq:gauss1p}, the amplitude of the envelope of the packet will decay due to diffraction into the transverse directions. Thus, in $3+1$ dimensions
 the particular measure of the superluminal signal, ${\textrm{Im}}(\psi(Y))$, decays 
as $Y^0= T$ -- the intervention time -- increases. To make this concrete, let us consider the Gaussian packet $|1_3\rangle$ peaked at $\mathbf k_0=(0,0,k_0)$ and calculate $\psi(Y)$ for arbitrary points in the $z-t$ plane, $Y=(t,0,0,z)$. Then $\psi(Y)$ can be evaluated in closed form~\cite[\S\,3.953]{gradshteyn2000table}:
\begin{equation}
\label{eq:psi4d}
\begin{array}{cl}
\psi(Y)
=\langle0|\hat\phi(t,0,0,z)|1_3\rangle&=(\pi\sigma^2)^{-\frac{3}4}\int \frac{d^3k}{(2\pi)^{\frac32}\sqrt{2|\mathbf k|}}e^{-\frac{(\mathbf k-\mathbf k_0)^2}{2\sigma^2}}e^{ik_\mu Y^\mu}\\
&=\frac{e^{- k_0^2/2\sigma^2}}{4\pi^{\frac34}}\frac{e^{v_-^2/4}D_{-\frac32}\left(v_-\right)-e^{v_+^2/4}D_{-\frac32}\left(v_+\right)}{k_0\sigma^{-2}+iz},
\end{array}
\end{equation}
where $v_\pm=i\sigma\left[t\pm (z- i k_0/\sigma^2)\right]$ and $D_\nu(z)$ is the parabolic cylinder function.

\begin{figure}
\includegraphics[width=0.75\textwidth]{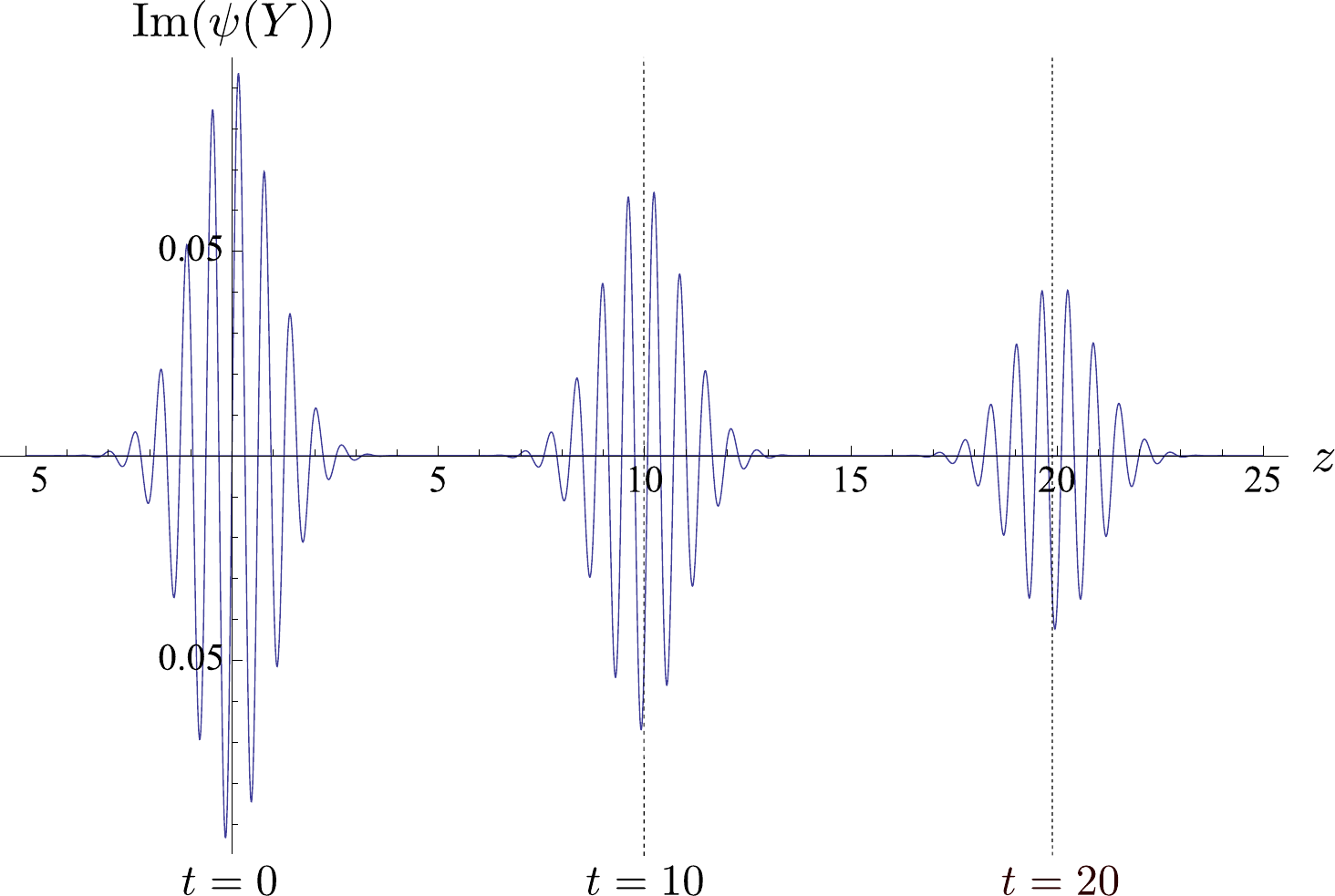}
\caption{The imaginary part of $\psi(t,0,0,z)$ for the Gaussian $1-$particle state with $\mathbf k_0=(0,0,10)$ and $\sigma=1$, plotted as a function of $z$ for different values of $t=0,10,20$.}
\label{fig:phi1particle}
\end{figure}

In Figure~\ref{fig:phi1particle}, 
the imaginary part of $\psi(t,0,0,z)$ for the Gaussian $1-$particle state with $\mathbf k_0=(0,0,10)$ and $\sigma=1$ is plotted as a function of $z$ for $t=0,10,20$. It can be seen that the amplitude of the 
packet envelope remains non-negligible for values of $t=Y^0$ which are
a few times the spatial width of the packet.  
If we choose $X$ to be at the origin, $X^\mu = (0,0,0,0)$, then at each of the 
representative times, points in the right hand half of the support of the packet are spacelike 
to $X$. Therefore $Y$ can be spacelike to $X$, in the support of the packet and such that  $Y^0 =$ few$\times$(spatial extent of packet)  and the measure of the superluminal 
signal $\textrm{Im}(\psi(Y))$ is non-negligible. Indeed, if we let $z=t+\delta$ with $0<\delta<1/\sigma$, such that $Y$ is spacelike to $X$ but still inside the support of the wavepacket, then the asymptotic expansion of~\eref{eq:psi4d} for large $t\gg k_0/\sigma^2$ gives~\cite{NISTdlmf}:
\begin{equation}
{\textrm{Im}}(\psi(Y))\sim \gamma \sqrt{k_0/\sigma}\cos(k_0\delta) t^{-1}\end{equation}
where $\gamma$ is a constant of order $\mathcal O(10^{-1})$. It follows that the envelope of the superluminal signal $\textrm{Im}(\psi(Y))$ falls off like $t^{-1}$.  
We conclude that one or more of the three interventions
-- the unitary kick at $X$, the ideal measurement of the
wavepacket between $t=0$ and $t=T=$few$\times$(spatial size of packet), 
or the measurement of $\hat\phi(Y)$ -- is impossible.
The measurement of the wavepacket is the most questionable
but to begin to address these questions properly we need to model the interventions
physically.

The same no-go results apply to unitary transformations:
an intervention by an external agent 
that produces a nonlocal unitary transformation on the state of the field, 
for example, a unitary transformation that rotates the state from one 1--particle state to an orthogonal 1--particle state, can
enable superluminal signalling. 
Consider
 the $2-$dimensional Hilbert space $\mathcal H$ spanned by two 1--particle states $|1\rangle$ and 
$| 1'\rangle$. This space is a subspace of the Fock space $\mathcal F=\mathcal H\oplus\mathcal H^\perp$ of the scalar field and $\mathcal H^\perp$ denotes its orthogonal complement. The field starts in a state $ |\psi\rangle = A|1\rangle
+ B |1'\rangle$. Consider the same 3 intervention regions, $\mathcal{O}_1$, 
$\mathcal{O}_2$ and $\mathcal{O}_3$, as before. 
A local unitary operation $e^{i \lambda \hat\phi(X)}$ is done at $X$ followed by
a unitary operation $U$ done in $\mathcal{O}_2$ where
\begin{equation}
\!U =e^{i\theta}( C |1\rangle\!\langle 1| + D  |1\rangle\!\langle 1'| - D^* |1'\rangle\!\langle 1| + C^* |1'\rangle\!\langle 1'| )+\mathbf 1^\perp\,,
\vphantom{\int}
\end{equation}
$|C|^2 + |D|^2 = 1$ and $\mathbf 1^\perp$ is the identity operator on $\mathcal H^\perp$. 
Finally, $\hat\phi(Y)$ is measured at $Y$. The expected value of the measurement of $\hat\phi(Y)$ is
\begin{equation}
\langle\hat\phi(Y)\rangle = \langle \psi| e^{- i \lambda \hat\phi(X)} U^\dagger \hat\phi(Y) U e^{ i \lambda \hat\phi(X)} | \psi \rangle\,.
\end{equation}
Differentiating w.r.t. $\lambda$ and setting $\lambda = 0$ we find  that the superluminal signal 
is nonzero if the following expression is nonzero:
\begin{equation}
2 {\textrm{Im}}  \langle \psi|  \hat\phi(X) U^\dagger \hat\phi(Y) U | \psi \rangle\,.
\label{eq:sulusirot}
\end{equation}
For $A=1$, $B=0$, $C=0$ and $D=-1$ we find
\begin{equation}
\langle \psi|  \hat\phi(X) U^\dagger \hat\phi(Y) U | \psi \rangle
= \psi(X)^* \psi'(Y) + \psi'(X)\psi(Y)^*
\label{eq:wavefuns}
\end{equation}
where $\psi(\xi) = \langle 0\,| \hat\phi(\xi) |\, 1\rangle$ and 
$\psi'(\xi) = \langle 0\,| \hat\phi(\xi) |\, 1'\rangle$ are the one particle 
wave functions. 

If the one particle states are single momentum states, this is easily worked out. Let us 
work in a box of side length $L$, so that
\begin{equation}
\hat\phi(X)=L^{-\frac{d}2}\sum_{\mathbf k}\frac1{\sqrt{2\omega_{\mathbf k}}}\left[a_{\mathbf{k}}^{\vphantom{\dagger}} u^{\vphantom{\dagger}}_{\mathbf k}(X) +a_{\mathbf{k}}^\dagger
 u^{*}_{\mathbf k}(X)\right] 
\end{equation}
where $u_{\mathbf k}(X)=e^{ik_\mu X^\mu}$. 
Let $|1\rangle=a_{\mathbf k}^\dagger|0\rangle$ and $|1'\rangle=a_{\mathbf k'}^\dagger|0\rangle$. Then the signal~\eref{eq:sulusirot} evaluates to
\begin{equation}
\!\!\!\!\!\!\frac{L^{-d}}{\sqrt{\omega_{\mathbf k}\omega_{\mathbf k'}}}\left[\sin(k'_\mu Y^\mu-k_\mu X^\mu)-\sin(k_\mu Y^\mu-k'_\mu X^\mu)\right].
\end{equation}
This can be nonzero for spacelike separated $X$ and $Y$. For example, for $\mathbf k=-\mathbf k' $ we obtain
\begin{equation}
-\frac{2L^{-d}}{\omega_{\mathbf k}}\sin\left[\omega_{\mathbf k}(Y^0-X^0)\right]\cos\left[\mathbf k\cdot(\mathbf Y+\mathbf X)\right],
\end{equation}
which is nonzero 
for choices of spacelike separated $Y$ and $X$ for which $Y^0-X^0$ is arbitrarily large. 
If  $|1\rangle$  and $|1'\rangle$ are wave packet states, the spreading of the packets becomes relevant. However,  similarly to the previous calculation, 
$X$ can be chosen in the support of $\psi$ at $t=0$ and $Y$ in the 
support of $\psi'$ at $t=T$ and there will be superluminal signalling for 
$T =$few$\times$(width of packet).
Unitary transformations that 
would apparently not violate relativistic causality are those which are 
products of unitaries that are perfectly localised at spacetime points on a 
spacelike hypersurface,
e.g. $e^{i \int d^dx f({\mathbf{x}}) \hat\phi(t, \mathbf{x})}$.

\section{Different rules}

The preceding observations show that modelling external
interventions on quantum fields as ideal measurements and unitary 
transformations in a straightforward generalisation of the text-book, 
operational rules for nonrelativistic quantum theory
fails. One response could be to put restrictions on the use of the formula \eref{eq:probs}
for the probabilities of outcomes of a collection of measurements. 

{\textit{Restricting the regions.}} One could require that the
relation  $\preceq$ between the intervention regions, $\mO_i$,  is already a partial order
{\textit{before}} the transitive closure is taken \cite{Sorkin:1993gg}.
 In the protocol considered above that would 
require  $X$ to be in the causal past of $Y$ and any signalling 
would then be causal by fiat. 
 This leaves
 the question, ``What is the probability of a certain sequence of outcomes of
 measurement-interventions in regions which are not partially ordered?'' without any answer,
though there seems to be no physical reason why 
the relative positions of the regions in spacetime should
affect agents' ability to 
do measurements in them.
Other, even more restrictive conditions on the regions $\mO_i$ might be considered. 
 For example, one might require that for every pair $(\mO_j, \mO_k)$ such that $j<k$,  either 
the two regions are entirely spacelike to each other or every point of $\mO_j$ 
is in the causal past of every point of $\mO_k$ \cite{Sorkin:1993gg}.\footnote{If one wanted to 
interpret this rule in terms of the quantum state it seems to mean that  
for each $\mO_i$ in turn, the 
quantum state collapses along the boundary of 
the future set $ \{y \,\vert \,  y\in J^+(x) \,\,\forall x \in \cup_{k=1}^i  \mO_k \}$.}
Once again, such rules would be silent on how to 
 make predictions when the interventions on 
 quantum fields take place in regions that do not satisfy these 
 conditions. 
 
{\textit{Restricting the observables.}} Alternatively one could 
leave the regions unrestricted and use the 
criterion that no superluminal signalling is allowed to
place conditions on the observables 
to which \eref{eq:probs} is to apply, the approach taken in \cite{PhysRevD.21.3316,PhysRevD.24.359,PhysRevD.34.1805,Aharonov:2008}. 
Beckman et al. \cite{Beckman:2001qs} comprehensively analyse 
the restriction that causality
places on  ``quantum operations'' -- including ideal 
measurements and unitary transformations as special cases --  on bipartite 
systems with tensor product Hilbert spaces. 

In quantum field theory, a particularly important example of a physical quantity 
that cannot be measured by an ideal measurement 
without causing superluminal signalling is a non-abelian Wilson 
loop \cite{Beckman:2001ck}. This seems to leave non-abelian gauge theory with no 
physical, gauge invariant  ``measureables'' at all! 
In scalar field theory, the situation seems not to be so dire. 
There is a class of observables which, should they be 
measurable by ideal measurements, would not causal superluminal 
signals: integrals over
spacelike hypersurfaces of local field operators and their 
conjugate momentum operators, such as those considered in 
\cite{Zych:2010yk}. At least, this is the conclusion one
would draw by analogy with the result for a bipartite, tensor product system
that an ideal measurement of a sum of local 
observables does not violate causality. For completeness, let us demonstrate this result.
Let $H = H_1\otimes H_2$ be a tensor product Hilbert space and let 
$A$ and $B$ be self-adjoint operators on $H_1$ and $H_2$ respectively. Suppose 
the system is initially in state $\rho_0$ and we consider the following protocol: 
a local unitary operation, $U_1$ on $H_1$, followed by a measurement of $X = A+B$. We
use the observation \cite{Sorkin:1993gg}  that the effect of a measurement of 
$X$ on the state is achieved by application of the unitary operator $e^{i \lambda X}$ with a 
random value of $\lambda$. $A$ and $B$ commute and so $e^{i \lambda X} = e^{i \lambda A}
e^{i \lambda B}$ which is a product of local unitaries which commute. 
Now consider the operation $U_1$ followed by application of $e^{i \lambda A}
e^{i \lambda B}$ followed by a partial trace over $H_1$, all on the state $\rho_0$. 
The reduced state for $H_2$ does not depend on $U_1$. 

This result suggests that in QFT no violation of relativistic causality 
would result from the ideal measurement of
 the (Heisenberg) observable 
$a_{\mathbf{k}}(t)  + a_{\mathbf{k}}^\dagger(t)
=  a_{\mathbf{k}} e^{-i \omega_{\mathbf{k}}t}  + a_{\mathbf{k}}^\dagger e^{i \omega_{\mathbf{k}}t}$, 
where $\omega_{\mathbf{k}}$ is the frequency for momentum $\mathbf{k}$, which may be written as
\begin{equation}
a_{\mathbf{k}}(t)  +  a_{\mathbf{k}}^\dagger(t) = \int d^{d}x \,\hat\phi(t, {\mathbf{x}}) F(\mathbf{x})
+ \int d^{d}x \,\hat\pi(t, {\mathbf{x}}) G(\mathbf{x})\label{observable}
\end{equation} 
where $\hat\pi$ is the canonical momentum of $\hat\phi$, and 
\begin{equation}
\begin{array}{ll}
F(\mathbf{x}) = (2 \omega_{\mathbf{k}})^{\frac{1}{2}}(2 \pi)^{-\frac{d}{2}} \cos \mathbf{k}\cdot\mathbf{x}\label{FF}\\
G(\mathbf{x}) = \left(\frac{2}{ \omega_{\mathbf{k}}}\right)^{\frac{1}{2}}(2 \pi)^{-\frac{d}{2}} \sin \mathbf{k}\cdot\mathbf{x}\label{GG}\,.
\end{array}
\end{equation} 
To make the single momentum states normalised by working in
a box of side length $L$, we could replace the $(2\pi)$ by $L$ in (\ref{FF}) and (\ref{GG}).
Although this observable
 is defined on the whole constant time hypersurface at $t$ it is a sum of 
truly local terms on the hypersurface. 
It is ``essentially local'' but not ``localised'' and therefore unlikely
 to be useful for QIP in Minkowski space. Localised 
smearings, such as those considered in \cite{Zych:2010yk}, have a better chance
of being useful. 

Another example of a more localised observable is $b(t)+ b^\dagger(t)$  where $b^\dagger(0)$ is 
defined by $b^\dagger(0) | 0\rangle = |1\rangle $, a one particle wave packet state.
If
\begin{equation}
b^\dagger(0) = \int d^d k \,\tilde\psi(\mathbf{k}) \,a^\dagger_{\mathbf{k}} 
\end{equation}
then the one particle wave function is
\begin{equation}
\psi(t, \mathbf{x}) = \int d^d k \,(2 \omega_{\mathbf{k}})^{-\frac{1}{2}}(2 \pi)^{-\frac{d}{2}} e^{i \mathbf{k}\cdot\mathbf{x}} e^{-i \omega_{\mathbf{k}} t} \tilde\psi(\mathbf{k})
\end{equation}
and
\begin{equation}
b(t) + b^\dagger(t) = \int d^{d}x\, \left(\hat\phi(t, {\mathbf{x}}) J(\mathbf{x})
+  \hat\pi(t, {\mathbf{x}}) K(\mathbf{x})\right) \label{observabletwo}
\end{equation} 
where 
\begin{equation}
\begin{array}{lcrlll}
J(\mathbf{x}) &= &
\int \frac{d^d k}{(2\pi)^{\frac{d}{2}}}
 \left(\frac{\omega_{\mathbf{k}}}{2}\right)^{\frac{1}{2}} 
 &\left[e^{i  \mathbf{k}\cdot\mathbf{x} } \tilde\psi(\mathbf{k}) + {\textrm{c.c.}}\right]
   \\
K(\mathbf{x}) &= &- i 
\int \frac{d^d k}{(2\pi)^{\frac{d}{2}}(2\omega_{\mathbf{k}})^{\frac{1}{2}}}
& \left[e^{i  \mathbf{k}\cdot\mathbf{x} }\tilde\psi(\mathbf{k}) - {\textrm{c.c.}}\right].
 \end{array}
\end{equation} 
$K(\mathbf{x})$ is (twice) the imaginary part of $\psi(0, \mathbf{x})  $ but 
$J(\mathbf{x})$ is not its real part. 

We see that for $b(t) + b^\dagger(t)$ to be localised, 
$\tilde\psi(\mathbf{k})$ must be such that 
$K$ and $J$ have bounded support in space. $K$ will do so if the 
one-particle wavefunction 
$\psi$
at $t=0$ does but requiring $J$ to have bounded support 
imposes an additional constraint and it is not clear that both constraints
can be exactly satisfied. If it is sufficient for $K$ and $J$ to be exponentially
small outside a bounded region of space then the Gaussian wave packet state 
(\ref{eq:gauss1p}) will do,  but, strictly, to measure 
$b(t) + b^\dagger(t)$ in this case requires an intervention region that includes
the whole spacelike hypersurface at $t$. If we accept that in this case
intervening in a bounded region can in principle result in an ideal measurement 
of $b(t) + b^\dagger(t)$ to a very good approximation, 
it is still an open question whether it can be measured
by an ideal measurement in practice. 

The
observables on which the algebraic approach to relativistic quantum field theory is based
are field operators smeared with test functions with bounded space{\textit{time}} support  
and Sorkin raised the question whether or not ideal measurements of such observables 
would enable superluminal signalling \cite{Sorkin:1993gg}. We note that for a free 
scalar field, since the field operators $\hat\phi(X)$ 
and $\hat\pi(X)$ at any spacetime point $X$ are linear 
combination of the field operators and the conjugate momentum operators on the
intersection of any spacelike hypersurface, $\Sigma$ and the causal past of $X$, such a spacetime
smearing of a free field is equal to a spatial smearing of local operators over $\Sigma$. This 
suggests that an ideal measurement of a spacetime smearing of a free field would not enable 
superluminal signalling. Of course, no field can be truly free, if only because measurements must be 
modelled by interactions, and for an interacting field, the question remains open.

\section{Detector models}
  
A more physical approach 
to addressing these questions -- and, for example, explaining which 
of the interventions 
considered in the first section is impossible -- 
is to construct explicit models of the interventions and measurements
as interactions between the field and other quantum systems. 
A quantum detector for example could act as the `qubit' that is manipulated 
 by the classical agent and
 the quantum field coupled to it would then be part of the quantum system that performs the
 processing task but would not be directly intervened upon by the external agent. 
A detector would have a 
 world-line or  world-volume in spacetime and 
 one would have to assume that ideal measurements and unitary transformations can 
 be performed on it by the external agent in some region.  
 
One prototype of such a detector is the 
Unruh-DeWitt (UD) detector \cite{Unruh:1976db,DeWitt:1980hx}. 
UD detectors and their ilk couple to the field operator itself and field operators at spacelike 
separated positions commute. It is an interesting question whether there is any observable
in the quantum field theory that can be said to be 
measured by an UD detector, but here we 
simply note that signalling between two UD detectors 
will  not occur {\textit{if}} their world-volumes are 
completely spacelike separated while they are switched on, 
 and (\ref{eq:probs}) is used for the probability of ideal 
measurements on the detectors themselves \cite{PhysRevA.81.012330}. 
Whether UD detectors are good models of 
realistic detectors that can be used in QIP is 
an open question.  In order to provide
a useful qubit a UD detector  needs to be switched
on and off at finite times and also itself prepared
and measured or manipulated at finite times. 
This requires the modelling of the switching 
process to be addressed with care \cite{Louko:2007mu}
but at least relativistic causality can be safe with UD detectors 
because they couple locally to the field. 

Are there other model detectors that might be applicable to
QIP in Minkowski spacetime? We briefly consider two that appear
in the literature. A detector that couples to one
single frequency mode of a scalar field in 1+1 dimensions is investigated in
\cite{MartinMartinez:2010sg,Dragan:2011hq}.   We
can imagine that the system is in a large box of size $L$ with periodic boundaries, say, so that the 
single mode state is normalisable. 
The Hamiltonian for the model is
\begin{equation}
\hat H(t) ={w}({\hat{d}}^\dagger {\hat{d}} + \frac{1}{2} )+\Omega ({\hat{f}}^\dagger {\hat{f}}+ \frac{1}{2})
+\lambda(t)(\hat{d}+ \hat{d}^\dagger)
\left(\hat{f} e^{i \Omega x} +\hat{f}^\dagger  e^{- i \Omega x} \right),
\end{equation}
where $x$ is the position of the detector which has an internal harmonic oscillator degree of 
freedom with frequency $w$ and raising operator $\hat{d}^\dagger$, and $\Omega$ is the 
frequency of the field mode with one-particle creation operator $\hat{f}^\dagger$. This causes superluminal signalling if interpreted as a model for a localised detector interacting
with a field in Minkowski spacetime, or in a box such that the light crossing time of
the box is longer than the timescale on which the model is valid. 

Consider, following  \cite{Dragan:2011hq},  the particular case of two static detectors at positions $x_i$,
with frequencies $w_i$ and couplings $\lambda_i(t)$, $i=1,2$. In order that the model 
be close to Minkowski spacetime, we assume that the box size $L$ is 
large compared to $|x_1 - x_2|$. In the interaction picture, the interaction Hamiltonians are
\begin{equation}
H_{i, {\textrm{int} }} ( t) 
= \lambda_i(t) (\hat{d}_i e^{- i \omega_i t} + \hat{d}_i^\dagger e^{i \omega_i t}) \left(\hat{f} e^{i \Omega(x_i - t)} +\hat{f}^\dagger  e^{- i \Omega(x_i - t)} \right)
\end{equation}
for $i = 1,2$. The time dependence in $\lambda_i(t)$ parametrizes the switching on and off of the detectors. 
The interaction Hamiltonians of the two 
detectors at spacelike separated positions do not commute. Indeed, 
defining $X_i^\mu := (t_i, x_i)$ and $\Omega^\mu := (\Omega, \Omega)$,  we have the
commutator 
\begin{equation}
\begin{array}{lcl}
[H_{1, {\textrm{int} }} ( t_1),  H_{2, {\textrm{int} }} ( t_2)]
&=& - 2 i \lambda_1(t_1)  \lambda_2(t_2) (\hat{d}_1 e^{- i \omega_1 t_1} + \hat{d}_1^\dagger e^{i \omega_1 t_1})\times \\
&&~~(\hat{d}_2 e^{- i \omega_2 t_2} + \hat{d}_2^\dagger e^{i \omega_2 t_2})\sin{\Omega_{\mu}(X_1 - X_2)^\mu}\,.
\end{array}
\end{equation}
This is nonzero for almost every pair of spacetime points along the trajectories of two 
detectors which are spacelike separated. 
This means that the unitary evolution operator in the interaction picture does not 
factorise into a product of an evolution operator for detector 1 and one for detector 2
and this leads to superluminal signalling between the detectors as we show explicitly
below. 

Let both detectors and the scalar field be initially in their free ground states at $t=0$ and assume that the detectors are switched off before $t=0$: $\lambda_i(t)=0$ for $t<0$. 
Working in the interaction picture, at time $T>0$ we measure for detector 1 the expectation value, denoted $E_1(T)$, of the operator
\begin{equation}
\mathcal{O}(T)= \omega_1(  \hat{d}_1(T)^\dag \hat{d}_1(T) + \frac{1}{2}) =\omega_1(\hat{d}_1^\dag \hat{d}_1+ \frac{1}{2}).
\end{equation}
$E_1(T)$ ought to be independent of $\lambda_2$ so long as detector 1 remains outside the causal future of the point $(0,x_2)$, i.e. when $T<|x_2-x_1|$. Otherwise, an observer at $(T,x_1)$ could use their detector to infer whether spacelike separated detector 2 has been switched on or not (i.e. whether $\lambda_2\neq0$ or not). We do not carefully model the switching process and
simply take $\lambda_i(t)$ to
be a step function, constant and nonzero only in the interval $0<t\leq T$
and then $E_1(T)$ can be evaluated explicitly using the Heisenberg Picture analysis presented in \cite{Dragan:2011hq}.  It can be shown that $E_1(T)$ does in general depend on $\lambda_2$, and by way of example let us evaluate it for a specific set of values: $w_1=w_2=\Omega=1$, $\lambda_1=\frac12$, $x_1=0$ and $x_2=2\pi$. With a constant coupling $\lambda_2(t)=\lambda_2$ for $0<t\leq T=\sqrt2\pi<|x_2-x_1|$ one obtains 
\begin{equation}
E_1(\sqrt{2}\pi)=\frac{1}{4} \left(2+\pi ^2\right)+\frac13\pi ^4\lambda_2^2+\mathcal O(\lambda_2^4).
\end{equation}
We see a second--order dependence on the coupling constant $\lambda_2$ of detector 2 -- a superluminal signal.

The signalling observed in the case of  finite mode coupling becomes more comprehensible if the detector is interpreted as a variant 
of an UD detector in which the detector couples both to the
field and to its conjugate momentum.
 Indeed, modifying the calculation that leads to (\ref{observable}), we 
find for the factor in the interaction hamiltonian $H_{1 ,{\textrm{int}}}(t)$ 
for the first detector
\begin{equation}
\hat{f} e^{i \Omega(x_1 - t)} +\hat{f}^\dagger  e^{- i \Omega(x_1 - t)} = \int dx \hat\phi(t, x) F_1(x)
+ \int dx \hat\pi(t, x) G_1(x)\label{newobservable}
\end{equation}
where
\begin{equation}
\begin{array}{lcl}
F_1(x) &=& (2 \Omega)^{\frac{1}{2}}L^{-\frac{1}{2}} \cos [{\Omega(x - x_1)}]\\
G_1(x) &=& \left(\frac{2}{ \Omega}\right)^{\frac{1}{2}}L^{-\frac{1}{2}} \sin[{\Omega(x-x_1)}]\,.
\end{array} 
\end{equation}
We can see that the effective spatial extension of the detector is not bounded: the 
detector is not localised and 
 two detectors which are nominally spacelike
separated actually overlap over all space.

In \cite{Costa:2008st} a more ambitious model of a detector 
defined using additional quantum fields 
coupled to the field being measured is considered.
Since the interaction between the fields is
local  no superluminal signalling can be enabled as a result of 
the interaction in and of itself. If taken literally,
the effective model 
of the detector given in equation (14)  in \cite{Costa:2008st} 
looks like it does enable superluminal signalling
since the interaction Hamiltonians of two detectors centred at spacelike separated 
positions do not commute. Just as in the calculation above, this will 
result in the response of one detector depending on whether or not there exists a
second one spacelike to it. Again,
this  happens because detectors constructed as described in \cite{Costa:2008st}
 cannot be genuinely spacelike separated: they may be centred around localised 
 positions but will always have some spatial overlap. The authors themselves 
state that the detector is nonlocal. However, 
 the question of whether the state of such a detector itself 
 is measurable by an ideal measurement immediately arises: does an 
 ideal measurement on the non-local detector enable superluminal signalling? 
This is not addressed in \cite{Costa:2008st} but it should not be hard to determine. 

\section{Cavities}
Those familiar with cavity QED might be surprised at these 
results. After all, in a cavity, ideal measurements of
observables such as particle number {\textit{can}} 
be done \cite{Johnson:2010} and there's a well-known, 
successful model of an atom-qubit interacting with QED in a cavity which is of the 
form investigated above in which a detector couples to a 
single mode of the field: the Jaynes-Cummings model (see {\textit{e.g}}
\cite{Miller:2005} for a review).  The reason there is no conflict with the 
results presented here is that the Jaynes-Cummings model is a 
 phenomenological model which applies only on time scales 
many orders of magnitude larger 
than the light crossing time of the cavity.
It does not describe the physics
on time scales of order the light crossing time or shorter\footnote{The laboratory values 
of the Rabi frequency and cavity size given in \cite{Miller:2005}
mean that the timescale for the model is about $10^5$ times the 
light crossing time of the cavity. Any result based on the assumption
that the dynamical evolution of the state of an atom in a cavity
is described by a Jaynes-Cummings type model on time scales much shorter than the 
Rabi time period, such as
\cite{MartinMartinez:2010sg} for example, 
is unphysical.}. If there are two atoms in a cavity, they cannot be placed
further apart than the size of the cavity and so their worldlines cannot remain spacelike to each other
on the time scales on which the 
 Jaynes-Cummings model applies and so no possibility of superluminal 
signalling arises in the model. 

The Jaynes-Cummings Hamiltonian and its relatives cannot 
model an atom-qubit coupled to a quantum field  
in Minkowski spacetime, or in any spacetime
where two atom-qubits can be placed at distances larger than 
 the timescale on which the detector model is valid.
 A model detector such as that 
considered in \cite{MartinMartinez:2010sg,Dragan:2011hq} would not do violence to  relativistic 
causality if it were interpreted phenomenologically 
and confined to a spacetime of much smaller spatial extent than the timescale on 
which the model is valid since this would effectively impose a non-relativistic causal order on the
intervention-events in the cavity. In the light of the phenomenological successes of such models,
one might be tempted to declare more widely that QFT is only physical in a box 
and that intervention regions $\mO_i$ 
must always have extent in time at least the light-crossing time of the box,
imposing a non-relativistic
 causal order on interventions on the field.  Such a rule would be
 highly restrictive and could not accommodate the questions that arise when the operations of the external agents on a quantum field take place in arbitrary, quasi-local
  spacetime regions, exactly the sort 
 of situation that a genuinely relativistic approach to QIP aims to describe.

{\section{Concluding Remarks}}

The questions remain. What are accurate models of physical
interventions on relativistic quantum fields in quasi-local regions of
Minkowski spacetime and what applications might they have in QIP?
Localised Unruh-DeWitt type detectors form one class of models: Are they realistic? Are there others?
The questions are interesting from the point of view of QIP but also because the attempt to answer them pushes us to address more foundational issues.
The struggle to describe measurement of relativistic quantum fields 
in a physical way reveals the limitations of the canonical and operational 
approach to quantum theory. Even when a detector is modelled as a 
quantum system, to be a useful qubit 
one still has to rely on the assumption that an external agent can 
measure {\textit{it}} by an ideal measurement and/or do unitary transformations on it. 
But a more fundamental description of the detector -- and indeed the agent -- 
would be in terms of quantum field theory and so one is not solving the 
problem but just pushing it one 
step away: how would one model the effect of an ideal measurement on the
detector if the detector were described within a quantum field theory 
({\textit{cf.}} \cite{Costa:2008st})? 

 A more physical approach is needed: we require a
framework for closed relativistic quantum systems including detectors, in which 
experimental, measurement-like situations can be analysed fully 
quantum mechanically in an essentially relativistic way. 

One proposal for such a framework takes the relativistic branch of
the `fork in the road' set out by Dirac early in the history of quantum mechanics. 
In the 1933 paper, ``The Lagrangian in quantum mechanics,''
Dirac wrote that the Lagrangian approach to 
classical mechanics is probably more fundamental than
the Hamiltonian approach because the former is relativistically 
invariant whereas the latter is ``essentially nonrelativistic'' \cite{Dirac:1933}.
In quantum theory the Hamiltonian approach 
leads to canonical 
quantisation, Hilbert space, operators, observables, transformation theory 
and the textbook rules for predicting the 
outcomes of measurements of observables 
and for the collapse of the state after an ideal measurement. 
These aspects of the canonical theory are indeed more or less
divorced from the spacetime nature of the physical world 
revealed by relativity. 

The relativistic alternative is to base quantum mechanics 
on the Lagrangian approach to classical mechanics and Dirac showed that 
 this leads to the path integral \cite{Dirac:1933}. The path integral 
 roots quantum theory firmly in spacetime
-- rather than Hilbert space -- as the arena for physics. 
In a path integral or 
sum-over-histories approach, the physical
world is described directly in terms of events in spacetime and Feynman's famous
paper on the path integral is aptly titled ``The spacetime approach to non-relativistic quantum mechanics''\cite{Feynman:1948}. 
 The path integral approach to the foundations of quantum theory 
has been championed in more recent times 
by Hartle \cite{Hartle:1991bb,
Hartle:1992as} and by Sorkin \cite{Sorkin:1994dt}. 
In \cite{Hartle:1993ip}  measurement situations in closed quantum systems are analysed
 from a sum-over-histories perspective, with an emphasis on a condition of decoherence. 
 In \cite{Sorkin:2010kg} a solution of the ``measurement problem'' of quantum mechanics is 
 set out which is valid, at least, when there are certain sorts of 
 permanent records of measurement outcomes. 
As a framework for  closed quantum systems which deals directly with 
spacetime events, the path integral approach is eminently suitable for 
the investigation of measurements on relativistic quantum fields in Minkowski spacetime. 

\vspace{0.1in}

\ack

We thank Andrzej Dragan, Jorma Louko, Alejandro Satz and Rafael Sorkin for useful and stimulating discussions. 
FD thanks Perimeter Institute for Theoretical Physics (Government of Canada through NSERC and by the Province of Ontario through MRI) and KITP, Santa Barbara (National Science Foundation under Grant No. NSF PHY11-25915)
for support. 
This research was supported in part by COST Action MP1006.
DMTB is supported by an EPSRC Studentship. The work of LB is
supported by an Imperial College Junior Research Fellowship.

\section*{References}



\end{document}